\documentclass[]{JHEP} 

\usepackage{epsfig}

\newcommand{\be}{\begin{equation}}   \newcommand{\ee}{\end{equation}}
\newcommand{\bear}{\begin{eqnarray}}
\newcommand{\eear}{\end{eqnarray}}
\newcommand{\ba}{\begin{array}}      \newcommand{\ea}{\end{array}}
\newcommand{\lsim}{\begin{array}{c}\,\sim\vspace{-21pt}\\< \end{array}}
\newcommand{\gsim}{\begin{array}{c}\,\sim\vspace{-21pt}\\> \end{array}}
\title{Light Axion within the Next-to-Minimal
       Supersymmetric Standard Model}

\author{Bogdan A.~Dobrescu\thanks{E-mail: bdob@fnal.gov} \ and 
        Konstantin T.~Matchev\thanks{E-mail: matchev@fnal.gov}\\
	Theoretical Physics Department\\
	Fermi National Accelerator Laboratory\\
        P.O. Box 500, Batavia, Illinois 60510, USA}
\received{\today} 		
\accepted{\today}		

\preprint{Fermilab-Pub-00/134-T \\ 
          hep-ph/0008192 \\ 
          August 17, 2000}	

\abstract{We analyze the Higgs sector in the Next-to-Minimal
Supersymmetric Standard Model, emphasizing the possibility of a
light CP-odd scalar (axion) in the spectrum.
We compute the coupling of the Standard-Model-like Higgs boson to a
pair of axions, and show that it can be large enough to modify 
the Higgs branching fractions, with a significant impact on
the Higgs searches. We delineate the range of parameters relevant
for this scenario, and also derive analytic expressions for the scalar 
masses and couplings in two special cases -- a decoupling limit where
all scalars other than the axion are heavier than the Standard-Model-like
Higgs boson, and the large $\tan\beta$ limit.}


\begin{document} 

\maketitle 

\section{Introduction}

In spite of the phenomenological
success of the Standard Model (SM), new physics is expected to appear
at some higher energy scale, and hopefully 
to provide a solution to the hierarchy problem,
the origin of fermion masses and CP-violation, and other
theoretical puzzles.
The new physics may change the properties of the Higgs 
boson, with a substantial impact for Higgs searches in collider
experiments. This is the case in extensions of the SM that 
include gauge singlet scalars, leading to Higgs 
boson decays into pairs of light neutral scalars
\cite{Dobrescu:2000jt}. 

Here we point out that the Higgs sector of the 
Next-to-Minimal Supersymmetric Standard Model (NMSSM)
\cite{Nilles:1983dy} includes an axion, {\it i.e.,} a 
pseudo-Nambu-Goldstone boson associated with an 
anomalous $U(1)$ symmetry, which for a range of parameters
is significantly lighter than the other scalars.
The NMSSM is a well motivated candidate for physics beyond 
the Standard Model. Not only does it provide a solution to the
hierarchy problem when combined with dynamical supersymmetry 
breaking, but it is also free of the $\mu$ problem
that plagues the minimal supersymmetric extension of the SM (MSSM).

In what follows we shall investigate in some detail
the Higgs spectrum and couplings in the NMSSM.
The purpose of our study is twofold. First, we seek to delineate 
the region of the NMSSM parameter space consistent with
the existing collider data which
exhibits a light axion. Second, for the
part of parameter space in question, we compute the
strength of the SM-like Higgs boson coupling to axion
pairs. If this coupling is sizable, it will have a profound
effect on the collider searches for the Higgs boson,
as the Higgs boson then decays mainly into light axion pairs,
and the $b\bar{b}$ signature is diluted.

The Higgs boson decay to axions persists even when the scale of 
supersymmetry is very large and the superpartners decouple.    
Furthermore, this phenomenon may occur in non-supersymmetric 
theories, {\it e.g.}, composite Higgs models 
\cite{Dobrescu:1999gv, Dobrescu:2000jt}
or Majoron models \cite{Shrock:1982kd}.

The plan of the paper is as follows.
In Section~\ref{nmssm} we introduce our notation,
derive the general tree-level mass matrices of the Higgs
sector, and list the trilinear couplings between one CP-even and
two CP-odd Higgs bosons in the NMSSM. In Section~\ref{raxion}
we concentrate on the case of a light axion, identifying
the relevant range of NMSSM parameters and discussing the
resulting masses and mixings in the CP-odd scalar sector.
We then derive simple analytic expressions for the Higgs
spectrum and couplings in two cases of interest --- 
a decoupling limit (Section~\ref{decoupling})
and large $\tan\beta$ (Section~\ref{largetb}).
A more generic case requires a numerical study, the  
results being presented in Section~\ref{numerical}.
Section~\ref{conclusions} is reserved for our conclusions.

\section{Next-to-Minimal Supersymmetric Standard Model}
\label{nmssm}

The Next-to-Minimal Supersymmetric Standard Model has the 
field content of the MSSM with the addition of a gauge-singlet 
chiral superfield\footnote{We shall use hatted symbols to represent
chiral superfields and symbols without hats for their
lowest (scalar) components.}, $\hat{S}$.
In addition to the usual Yukawa-type couplings of the Higgs
superfields, $\hat{H}_u$ and $\hat{H}_d$, to the 
three generations of quark and lepton superfields,
the superpotential $W$ also includes the following terms
involving $\hat{S}$:
\be
W = \lambda \hat{H}_u \hat{H}_d \hat{S} + \frac{\kappa}{3} \hat{S}^3 ~.
\label{superpotential}
\ee
We assume that the $R$-parity violating terms involving quarks and leptons, 
and the dimensionful $\hat{H}_u \hat{H}_d$, $\hat{S}^2$ and $\hat{S}$ 
terms are forbidden by a gauge symmetry which is spontaneously broken above 
the TeV scale \cite{Cheng:1998nb}.
A sector of dynamical supersymmetry breaking is supposed to induce masses 
for the squarks, sleptons and gauginos, as well as soft supersymmetry 
breaking terms involving the scalar components of $\hat{H}_u$, $\hat{H}_d$
and $\hat{S}$:
\be
V_{\rm soft} = M_{H_u}^2 |H_u|^2 + M_{H_d}^2 |H_d|^2 +
M_S^2 |S|^2 + \sqrt{2}
\left(m_\lambda H_u^\top i\sigma_2 H_d S - \frac{m_\kappa}{3} S^3 
+ {\rm h.c.} \right)\, .
\ee
The $H_u H_d$, $S^2$ and $S$ soft terms are forbidden by
the same symmetry which prevents the $\hat{H}_u \hat{H}_d$,
$\hat{S}^2$ and $\hat{S}$ terms in the superpotential.

In what follows we treat the NMSSM as a low-energy effective field
theory valid below some scale, say in the TeV range. 
Therefore, we define the five mass parameters from $V_{\rm soft}$
as free parameters at the electroweak scale, $v \approx 246$ GeV.
Phenomenologically, they are constrained by the requirement of having an
electroweak asymmetric vacuum and a spectrum of scalars heavier than the 
current experimental bounds.

The scalar potential for the Higgs sector of the NMSSM is 
\be
V = \left| \lambda H_u^\top i\sigma_2 H_d + \kappa S^2 \right|^2 
+ \lambda^2\left( |H_u|^2 + |H_d|^2 \right) |S|^2 
+ V_D + V_{\rm soft} ~,
\label{potential}
\ee
with the usual $D$-term contributions
\be
V_D = \frac{M_Z^2}{2v^2} \left( |H_u|^2 - |H_d|^2 \right)^2 
+ 2 \frac{M_W^2}{v^2} \left|H_u^\dagger H_d \right|^2 ~,
\label{d-potential}
\ee
where $M_W$ ($M_Z$) is the $W$-boson ($Z$-boson) mass.
If $V$ has a minimum where the vacuum expectation values
(VEVs) of $H_u$ and $H_d$ are aligned and non-zero,
the electroweak symmetry is spontaneously broken.
Then, in addition to the longitudinal $W$ and $Z$
({\it i.e.} the Nambu-Goldstone modes $G^+$ and $G^0$, respectively)
the scalar spectrum includes a charged Higgs boson, $H^\pm$,
and five neutral states.
In general all five neutral states mix. However, if 
${\rm Im} (\lambda \kappa^* m_\lambda m_\kappa^*) \ll v^2$,
then the scalar potential is approximately CP invariant.
We will assume that this is the case,
so that the mixing of the CP-even neutral scalars
with the CP-odd ones can be ignored.
It is convenient to derive the spectrum using the basis where
the CP-even, $h^0_v, H_v^0, h^0_s$, and the CP-odd, $A^0_v, A^0_s$,
states are defined as follows:
\bear
H_d & = & \left( \begin{array}{c}
\frac{1}{\sqrt{2}} 
\left[ (v + h^0_v - i G^0  )\cos\beta 
     - (    H^0_v - i A^0_v)\sin\beta \right] \\ [4mm]
                     -\ G^- \cos\beta 
                      + H^- \sin\beta 
\end{array} \right) \; ~,
\nonumber \\ [7mm]
H_u & = &\left( \begin{array}{c} 
                        G^+ \sin\beta 
                      + H^+ \cos\beta  \\ [4mm]
\frac{1}{\sqrt{2}} 
\left[ (v + h^0_v + i G^0  )\sin\beta 
     + (    H^0_v + i A^0_v)\cos\beta
\right]
\end{array} \right) \; ~,
\nonumber \\ [6mm]
S & = & \frac{1}{\sqrt{2}} \left( s + h^0_s + i A^0_s \right)\; ~,
\label{vacuum}
\eear
with $\tan\beta\equiv \langle H_u^2 \rangle / \langle H_d^1 \rangle$.
Notice that in order to generate masses for both up-type and
down-type quarks one needs VEVs for both $H_u$ and $H_d$. 
The $m_\lambda H_u^\top i\sigma_2 H_d S$ soft term forces $S$ to have a 
non-zero VEV, $s$. We choose $0 < \beta < \pi$, which requires 
$s>0$ in order to minimize the first term in Eq.~(\ref{potential}).

There are several advantages of using the basis (\ref{vacuum})
from the beginning. First, notice that $h^0_v$ is rotated in the same
way as $v$, and is exactly the linear combination of
$H_d^1$ and $H_u^2$ responsible for the masses of the $W$ and $Z$
gauge bosons. Consequently, $h^0_v$ is also the state which
has trilinear couplings to $W$ and $Z$ pairs, and can be produced
in association with a $W$ or $Z$ at the Tevatron or LEP.
In short, $h^0_v$ can be identified with the SM-like Higgs boson,
$h^0$. In addition, the Nambu-Goldstone modes $G^\pm$ and $G^0$
decouple from the corresponding mass matrices and need not be
considered in our further analysis. (The basis (\ref{vacuum})
was considered also in \cite{Yeghian:1999kr};
for results in the more conventional basis see, {\it e.g.} 
\cite{Franke:1997tc}.)

The extremization conditions for the scalar potential allow us to 
replace the three mass-squared parameters from $V_{\rm soft}$ 
by the three VEVs, $v\sin\beta, \, v\cos\beta, \, s$:
\bear
M_{H_d}^2 & = & -\frac{\lambda^2}{2} \left(s^2 + v^2 \sin^2\!\beta\right)
+ \frac{\lambda\kappa}{2} s^2\tan\beta -\frac{M_Z^2}{2} \cos 2\beta
+ m_\lambda s\tan\beta~,
\nonumber \\ [3mm]
M_{H_u}^2 & = & -\frac{\lambda^2}{2} \left(s^2 + v^2 \cos^2\!\beta\right)
+ \frac{\lambda\kappa s^2}{2\tan\beta} + \frac{M_Z^2}{2} \cos 2\beta
+ \frac{m_\lambda s}{\tan\beta}~,
\nonumber \\ [3mm]
M_S^2 & = & -\frac{\lambda^2}{2} v^2 
+ \frac{\lambda\kappa}{2} v^2 \sin 2\beta - \kappa^2 s^2
+ \frac{m_\lambda v^2}{2 s} \sin 2\beta
+ m_\kappa s ~.
\eear
Therefore, the scalar masses depend on the following six
unknown parameters: $\tan\beta$, $\lambda, \,\kappa$, $s$, $m_\lambda, \,
m_\kappa$.

The squared-mass matrix for the CP-even scalars, $h^0_v, H_v^0, h^0_s$,
is given by 
\be
 {\cal M}^2_h = v^2 \, \left( \begin{array}{ccc}  
r + \frac{\textstyle M_Z^2}{\textstyle v^2}
& r \cot 2\beta
& \lambda^2\frac{\textstyle s}{\textstyle v} - R
 \\ [4mm] 
r \cot 2\beta
& \;\; - r 
+ \frac{\textstyle \lambda\kappa s^2 + 
2 m_\lambda s}{\textstyle v^2 \sin 2\beta} \;\;
& - R \cot 2\beta
\\ [4mm] 
\lambda^2\frac{\textstyle s}{\textstyle v} - R
& - R \cot 2\beta
& \frac{\textstyle s}{\textstyle v^2}
\left(2 \kappa^2 s - m_\kappa \right)
+ \frac{\textstyle m_\lambda}{\textstyle 2 s}\sin 2\beta
 \end{array} \right) 
\label{cp-even}
\ee
where we have defined
\be
r \equiv
\left( \frac{\lambda^2}{2} - \frac{ M_Z^2}{v^2}\right)
\sin^2\! 2\beta ~,
\label{smallr}
\ee
and
\be
R \equiv \frac{1}{v} 
\left(\lambda\kappa s + m_\lambda\right) \sin 2\beta ~.
\label{bigr}
\ee
We label the CP-even mass eigenstates in order of increasing mass,
\be 
\left( \begin{array}{c} H^0_1 \\ H^0_2 \\ H^0_3 \end{array} \right)
= U \left( \begin{array}{c} h^0_v \\ H_v^0 \\ h_s^0 \end{array} \right) ~,
\label{evenrotation}
\ee
where the $3\times3$ orthogonal matrix $U$ 
may be obtained by diagonalizing ${\cal M}^2_h$. 
Whenever there is no confusion,
we shall use an alternative labelling of the Higgs
boson mass eigenstates, in analogy to the MSSM.
We shall use $h^0$ for the SM-like mass state, {\it i.e.}
the one with the largest projection onto $h^0_v$;
$H^0$ for the state corresponding to the ``heavy''
CP-even Higgs boson of the MSSM, {\it i.e.}
the one with the largest projection onto $H^0_v$;
and $H'^0$ for the state corresponding to the
additional singlet of the NMSSM, {\it i.e.}
the one with the largest projection onto $h^0_s$.
As can be readily seen from (\ref{cp-even}), in the limit
of large $s$, the SM-like Higgs
boson $h^0$ is identified with $H^0_1$.
However, for small values of $s$, 
$h^0$ can also be $H^0_2$ or even $H^0_3$,
depending on the other parameters.

The CP-odd states, $A_v^0, A_s^0$ have the following  
squared-mass matrix:
\be
{\cal M}^2_A = 
\left( \begin{array}{cc} \frac{\textstyle \lambda\kappa s^2 
+ 2 m_\lambda s }{\textstyle \sin 2\beta}
&  - v\left(\lambda\kappa s - m_\lambda\right)
 \\ [3mm] 
- v\left(\lambda\kappa s - m_\lambda\right) 
& \;\;  \left(\lambda\kappa 
+ \frac{\textstyle m_\lambda}{\textstyle 2 s} \right) v^2 \sin 2\beta
+ 3 s m_\kappa
 \end{array} \right) ~.
\label{cp-odd}
\ee
The CP-odd mass eigenstates may be written in terms of the $A_v^0$ and 
$A_s^0$ states:
\be 
\left( \begin{array}{c} A^0_1 \\ A^0_2 \end{array} \right)
= 
\left( \begin{array}{cc} \cos\theta_A &\quad   \sin\theta_A \\
   -\sin\theta_A &\quad   \cos\theta_A  \end{array} \right)
\left( \begin{array}{c} A^0_v \\ A^0_s \end{array} \right) ~,
\label{oddrotation}
\ee
with a mixing angle $\theta_A$ that satisfies
\be
\tan 2\theta_A = \frac{-4vs(\lambda\kappa s - m_\lambda) \sin 2\beta}
{v^2 \sin^2\!2\beta(2\lambda\kappa s + m_\lambda)
-2s^2 (\lambda\kappa s + 2m_\lambda -3 m_\kappa \sin 2\beta )} ~.
\label{tant}
\ee

Finally, the charged Higgs boson has a mass
\be
M^2_{H^\pm} =  \frac{\textstyle \lambda\kappa s^2 
+ 2 m_\lambda s}{\textstyle \sin 2\beta} - \frac{\lambda^2}{2} v^2
+ M_W^2 ~.
\label{mcharged}
\ee
The vacuum defined by Eq.~(\ref{vacuum}) is indeed a viable minimum 
of the scalar potential provided all physical scalars have 
positive masses. Therefore, all eigenvalues of 
${\cal M}^2_h$ and ${\cal M}^2_A$ have to be positive, and 
$M^2_{H^\pm} > 0$. We will analyze the constraints 
on the  parameter space imposed by these conditions
both numerically (Section~\ref{numerical}) and 
analytically in certain interesting limits
(Sections~\ref{decoupling} and \ref{largetb}).
 
In particular, we will be concentrating on the case where
one of the CP-odd scalars is light, and therefore the
neutral CP-even scalars may decay into a pair of CP-odd states.
The relevant trilinear couplings in the basis (\ref{vacuum})
are given at tree-level by 
\bear 
{\cal L}_{HAA} &=& - \frac{v}{2}\left\{ \, \left[
\left(\frac{\lambda^2}{2} + r \cot^2\!2\beta \right) h_v^0
- r \cot 2\beta \, H_v^0 + \left(R + \lambda^2\frac{s}{v}  \right)
h_s^0 \right] (A_v^0)^2  \right.
\nonumber \\ [2mm]
&& \qquad + \left[ \lambda \left( \kappa \sin 2\beta + \lambda \right) h_v^0
+ \lambda\kappa \cos 2\beta \, H_v^0 
+ 2 \frac{\kappa^2 s - m_\kappa}{v}  h_s^0 \right] (A_s^0)^2 
\nonumber \\ [2mm]
&& \qquad \left. - 2 \left( \frac{\lambda\kappa s - m_\lambda}{v} h_v^0 +
\lambda\kappa h_s^0 \right) A_vA_s \right\} ~.
\label{haa}
\eear
In order to compute the Higgs decay width into CP-odd scalars, 
one has first to determine the rotation matrix $U$ and the mixing angle 
$\theta_A$, and then to derive the trilinear couplings in the mass 
eigenstate basis. We will perform this computation in the following 
sections.

\section{The Case of a Light Axion}
\label{raxion}

\subsection{Approximate $R$-symmetry}

The scalar potential $V$ has no global continuous symmetry. 
However, in the limit where the coefficients of the 
trilinear terms vanish, $m_\lambda, m_\kappa \rightarrow 0$,
the potential has a global $U(1)_R$ symmetry under which the
$S$ charge, $y_S \neq 0$, is half the charge of $H_uH_d$.
This symmetry is spontaneously broken by the VEVs of $H_u, H_d$ and $S$,
so that apparently there is a Nambu-Goldstone boson in the spectrum. 
In addition, $U(1)_R$ is explicitly broken by the QCD anomaly. 
To see this, note that the  Yukawa terms responsible for quark 
masses impose constraints on the $U(1)_R$ charges of the quarks
such that the  $[SU(3)_C]^2 \times U(1)_R$ anomaly is proportional
to $y_S$. Hence the Nambu-Goldstone boson is in fact an axion, 
and there is a small contribution to its mass from QCD.

Furthermore, there is another source of explicit $U(1)_R$ breaking.
To see this, note that the form of the superpotential
(\ref{superpotential})
requires that $U(1)_R$ does not commute with supersymmetry, 
{\it i.e.} the fermion components of the $\hat{H}_u,\hat{H}_d$
and $\hat{S}$ superfields have different $U(1)_R$ charges
than the corresponding scalars.
Therefore, this $U(1)_R$ is an $R$-symmetry.
Given that the gauginos carry $R$-charge, it follows that 
$U(1)_R$ is explicitly broken by the gaugino masses.
This effect appears in the effective potential via one-loop diagrams
with $H_{u,d}$-gaugino-Higgsino vertices.
Although this contribution to the axion mass 
is larger than the contribution from the anomaly, 
the loop suppression implies that the axion is 
lighter than the other scalars by more than an order of magnitude
in the limit $m_\lambda, m_\kappa \rightarrow 0$.

One has to be alert for potential confusions regarding the ``axion''
label used in this paper. This axion is not useful for solving the
strong CP problem because the explicit $U(1)_R$ breaking due to 
gaugino masses exceeds the anomaly contribution. Also, the axion
associated with this approximate $U(1)_R$ is different from
the Peccei-Quinn axion associated with the global $U(1)$ 
recovered in the $\kappa, m_\kappa \rightarrow 0$ limit.

Another confusion may be caused by the 
$R$-axion from the dynamical supersymmetry breaking sector.
Typically, the models of dynamical supersymmetry breaking have 
a spontaneously broken $R$-symmetry \cite{Nelson:1994nf}. 
The associated Nambu-Goldstone boson is called an
$R$-axion and would be massless in the absence of a source of 
$R$-symmetry breaking, such as a term in the superpotential required 
for cancelling the cosmological constant \cite{Bagger:1994hh}.
If there is a hidden sector where  supersymmetry is dynamically
broken, and supersymmetry breaking is mediated from this sector 
to the NMSSM via supergravity, gauge interactions, or any other
mechanism, then there is mixing between the $R$-axion and the 
axion discussed in this paper. However, this mixing is suppressed
by the scale associated with supersymmetry breaking mediation, and
may be ignored for practical purposes.
It is important however, that the 
spontaneous breaking of the $R$-symmetry within the hidden sector 
is the source of gaugino masses in the NMSSM. Therefore, 
the existence of the $R$-axion in the hidden sector requires
a mass for the $U(1)_R$ axion from the NMSSM.

\subsection{Properties of the axion}

The light axion may also be identified by
studying the spectrum of CP-odd states given in Section 2.
In what follows we will expand in $m_{\lambda}/v$ and
$m_{\kappa}/v$,
neglecting the loop effects, which is appropriate for
$1 \gg m_{\lambda,\kappa}/v \gsim {\cal O}(10^{-3})$.
The lightest CP-odd neutral scalar, $A_1^0$, is the axion 
associated with the approximate $U(1)_R$ symmetry, and its mass,
\be
M_{A_1} = \sqrt{3s} \left( m_\kappa \sin^2\!\theta_A 
+ \frac{3m_\lambda\cos^2\!\theta_A}{2\sin 2\beta}\right)^{\! 1/2} 
+ {\cal O}(m_{\lambda,\kappa}^{3/2}/\sqrt{v}) ~,
\label{mA1}
\ee
vanishes in the limit $m_\lambda, m_\kappa \rightarrow 0$,
in agreement with the arguments presented above.
The other CP-odd neutral scalar, $A_2^0$, has a mass
\be
M_{A_2} \approx \frac{v}{\cos\theta_A}\sqrt{\lambda\kappa\sin 2\beta} 
+ {\cal O}(m_\lambda,m_\kappa) ~.
\ee
The mixing angle $0< \theta_A < \pi/2$ also has a simple form:
\be
\tan\theta_A = \frac{s}{v \sin 2\beta} + {\cal O}(m_{\lambda,\kappa}/v) ~.
\label{tantheta}
\ee
Notice that in the small $m_\lambda$, $m_\kappa$ limit
the axion mass $M_{A_1}$ depends on only four out of the six
input parameters of the Higgs sector: $s, \tan\beta, m_\lambda$
and $m_\kappa$. The dependence on $\lambda$ and $\kappa$
drops out because they are not related to the breaking of
the $U(1)_R$ symmetry.

The couplings of the axion to quarks and leptons may easily be
derived by observing that its $A^0_v$ component has the couplings of the 
MSSM CP-odd scalar while $A^0_s$ does not  couple
to the quarks and leptons. Therefore, the couplings are proportional
to the fermion masses within the up- and down-type sectors separately: 
\be
\frac{\cos\theta_A}{v} \left(m_u \cot\beta \; \bar{u}\gamma_5 u 
+ m_d \tan\beta \;\bar{d}\gamma_5 d \right) i A_1^0 ~.
\label{A0couplings}
\ee
However, in contrast to the MSSM CP-odd scalar,
the couplings of $A_1^0$ to down-type
fermions are {\em not} enhanced by $\tan\beta$ because 
in the large
$\tan\beta$ limit Eq.~(\ref{tantheta}) gives
\be
\cos\theta_A \approx {v\over s}{2\over\tan\beta}~.
\label{costhetaA}
\ee
We then see that the $\tan\beta$ enhancement of the $A^0_v$ coupling
to down-type fermions is exactly compensated by a $\tan\beta$
suppression in the mixing angle $\theta_A$. On the other hand,
the $A^0_1$ couplings to up-type fermions are doubly suppressed
by $\tan\beta$, and Eq.~(\ref{A0couplings}) becomes
\be
\frac{2}{s} \left({m_u \over \tan^2\beta} \; \bar{u}\gamma_5 u 
+ m_d \;\bar{d}\gamma_5 d \right) i A_1^0 ~.
\ee
The phenomenological implications of this result are clear:
when  $\tan\beta \gg 1$
the cross-sections for $A^0_1$ production in association with a pair
of down-type fermions ({\em e.g.} $b\bar{b}$) do not depend on
$\tan\beta$, 
while the $A^0_1$ branching ratios into down-type fermions are
enhanced.

Of special interest for Higgs phenomenology is the axion
coupling to the SM-like Higgs boson $h^0$,
\be 
{\cal L}_{h^0A_1^0A_1^0} = \frac{c}{2} v \ h^0 A_1^0A_1^0 ~,
\label{trilinear}
\ee
where 
\bear
c &=&  
\frac{-1}{1 + \frac{v^2}{s^2} \sin^2\! 2\beta}\left\{ 
U_{i1} \left[ \lambda \left(\lambda - \kappa \sin 2\beta \right) 
+ \frac{v^2}{s^2}\left( \frac{\lambda^2}{2} \sin^2\! 2\beta 
+ r \cos^2\! 2\beta \right)\right] \right.
\nonumber \\ [2mm]
&&\qquad\qquad\qquad + \left.  U_{i2} \cos 2\beta 
\left( \lambda\kappa - \frac{v^2 }{s^2} r \sin 2\beta \right)
\right. \nonumber \\ [2mm]
&&\qquad\qquad\qquad
+ \left. U_{i3} \frac{s}{v}  \left[ 2\kappa^2
+ \frac{v^2}{s^2} \lambda \sin 2\beta 
\left( \kappa \sin^2\! 2\beta - 2 \kappa
+ \lambda \sin 2\beta\right)\right]\right\} .
\label{haarot}
\eear
Here the index $i$ labels the mass eigenstate corresponding to $h^0$.
Although analytical expressions for the elements of the rotation
matrix $U$ may be written in general \cite{Han:2000jc}, here
we prefer to derive the trilinear coupling $c$ as a series in $v/s$
(Section~\ref{decoupling}) or $1/\tan\beta$ (Section~\ref{largetb}),
which should help the reader gain some insight into the
numerical results of Section~\ref{numerical}.

\section{The Decoupling Limit, $ s \gg v$}
\label{decoupling}

In this Section we study the decoupling limit of the NMSSM
in which $A^0_1$ and the lightest CP-even scalar, $H^0_1$, are
much lighter than the other scalars. This situation arises
when the gauge-singlet VEV is large\footnote{Note that
the dimensionless couplings $\lambda$ and $\kappa$
from the superpotential of the Higgs sector are usually
assumed to be roughly of order one, {\it i.e.} we 
ignore cases where a hierarchy is generated 
through some kind of fine-tuning of the couplings.}, $s \gg v$,
as can be seen by inspecting the expressions for
${\cal M}^2_h$, ${\cal M}^2_A$ and $M_{H^\pm}^2$
given in Section~\ref{nmssm}. Under those circumstances,
$H^0_1$ is simply the SM-like Higgs boson $h^0$,
and the low-energy limit of the model is just the SM,
with the addition of one (mostly singlet) CP-odd scalar, $A^0_1$.

In order to analyze the CP-even neutral states we
diagonalize ${\cal M}^2_h$ by expanding in $v/s$. We keep only the 
leading order in $m_{\lambda,\kappa}/v$ because we will only be 
interested in the region of parameter space where there is a light axion.
The CP-even mass eigenstates are given by Eq.~(\ref{evenrotation}),
with
\bear
U_{11} &=& 1 - \frac{v^2}{s^2} \frac{\lambda^2}{8\kappa^4}
\left(\lambda - \kappa \sin 2\beta \right)^2 
+ {\cal O}\left({v^4\over s^4}\right)~,  \nonumber \\ [3mm]
U_{12} &=& - {v^2\over s^2} {\lambda^2\over 4\kappa^2}
\left( 1 - {2\kappa\over\lambda^3}
{M_Z^2\over v^2}\sin 2 \beta \right)\sin 4\beta  
+ {\cal O}\left({v^3\over s^3}\right)~,  \nonumber \\ [3mm]
U_{13}&=&-\frac{v}{s} 
          \frac{\lambda^2}{2\kappa^2} 
          \left(1-{\kappa\over\lambda}\sin2\beta\right) 
- \frac{v^3}{s^3} 
\left\{ \frac{M_Z^2}{v^2}
        \frac{\cos^2\!2\beta}{4\kappa^4}
        (\lambda + \kappa \sin 2\beta ) 
        (\lambda - 2 \kappa \sin 2\beta) 
         \right. \nonumber  \\ [3mm]
&&-\left. \frac{\lambda^3}{4\kappa^3} \left[
          \frac{\lambda-\kappa\sin 2\beta }{4\kappa^3} 
          \left(3\lambda^2
               -6\lambda\kappa\sin 2\beta
               +\kappa^2\sin^2\!2\beta \right)  
         - \sin 2\beta\cos^2\!2\beta\right]
\right\} \nonumber  \\ [3mm]
&&+\ {\cal O}\left({v^4\over s^4}\right)\, . \label{us}
\eear
The other elements of the orthogonal matrix $U$, 
\bear
U_{21}&=& - {v^2\over s^2} \sin^2\! 2 \beta \cos 2 \beta
     \left[{\lambda^2\over2\kappa(\lambda-2\kappa\sin2\beta)}
          +{M_Z^2\over\lambda\kappa v^2}\right]	
     + {\cal O}\left({v^3\over s^3}\right)~,
\nonumber \\ [3mm]
U_{31}&=& \frac{v}{s} 
          \frac{\lambda^2}{2\kappa^2} 
          \left(1-{\kappa\over\lambda}\sin2\beta\right) 
       + {\cal O}\left({v^3\over s^3}\right)~, 
\nonumber \\ [3mm]
U_{23}&=& -\ U_{32}
       = - {v\over s}\,
            {\lambda\sin2\beta\cos2\beta \over
            \lambda-2\kappa\sin2\beta}
       + {\cal O}\left({v^3\over s^3}\right)~,
\nonumber \\ [3mm]
U_{22}&=&  1 
       - {v^2\over 2s^2}\,
       \left({\lambda\sin2\beta\cos2\beta\over
             \lambda-2\kappa\sin2\beta}\right)^2
       + {\cal O}\left({v^4\over s^4}\right)~,
\nonumber \\ [3mm]
U_{33}&=&  1 
       - {v^2\over 2s^2}\,
       \left\{\left({\lambda\sin2\beta\cos2\beta\over
                    \lambda-2\kappa\sin2\beta}\right)^2
            +\left[{\lambda(\lambda-\kappa\sin2\beta)\over
                    2\kappa^2}\right]^2
       \right\}
       + {\cal O}\left({v^4\over s^4}\right)~,
\eear
are less important in what follows, and we only list them
here for completeness. Notice the $s/v$ enhancement in the
last line of Eq.~(\ref{haarot}), which requires us to compute
$U_{13}$ up to an additional order in the $v/s$ expansion.

After computing the eigenvalues of ${\cal M}^2_h$ as a power
series in $v/s$, we find the following CP-even scalar masses:
\bear
M_{h^0}  &=&  v \left[ 
             \frac{M_Z^2}{v^2} \cos^2\! 2\beta 
           - \frac{\lambda^3}{2\kappa^2}
             \left(\lambda - 2\kappa\sin 2\beta \right)  
                \right]^{\! 1/2} 
          + {\cal O}\left({v^3\over s^2}\right) ~,
             \label{mhs} \\ [4mm]
M_{H^0} &=&  s \left(
             \frac{\lambda\kappa}{\sin 2\beta}\right)^{1/2} 
          + {\cal O}\left({v^2\over s}\right) ~, 
               \label{mHapprox}  \\ [3mm]
M_{H'^0}&=& \sqrt{2} \kappa s + {\cal O}\left({v^2\over s}\right) ~.
\label{mHsapprox}
\eear
We see from Eqs.~(\ref{mHapprox}), (\ref{cp-odd}) and (\ref{mcharged})
that to leading order in $v/s$, the $H^0$, $A^0_2$ and $H^\pm$
scalars are degenerate, forming a weak-doublet complex scalar of mass
\be
s \left({\lambda\kappa\over\sin 2\beta}\right)^{1/2} \gg v~.
\ee
These are the familiar ``heavy'' Higgs bosons of the MSSM.
Eq.~(\ref{mHsapprox}) confirms that the $H'^0$ scalar is also
heavy, with a mass of the order of the gauge singlet VEV $s$.
Therefore, in the limit $s\gg v$ considered in this section,
the only surviving scalars with masses of order the electroweak
scale or smaller are the SM-like Higgs boson $h^0$
and the axion $A^0_1$.

\FIGURE[t]{\epsfig{file=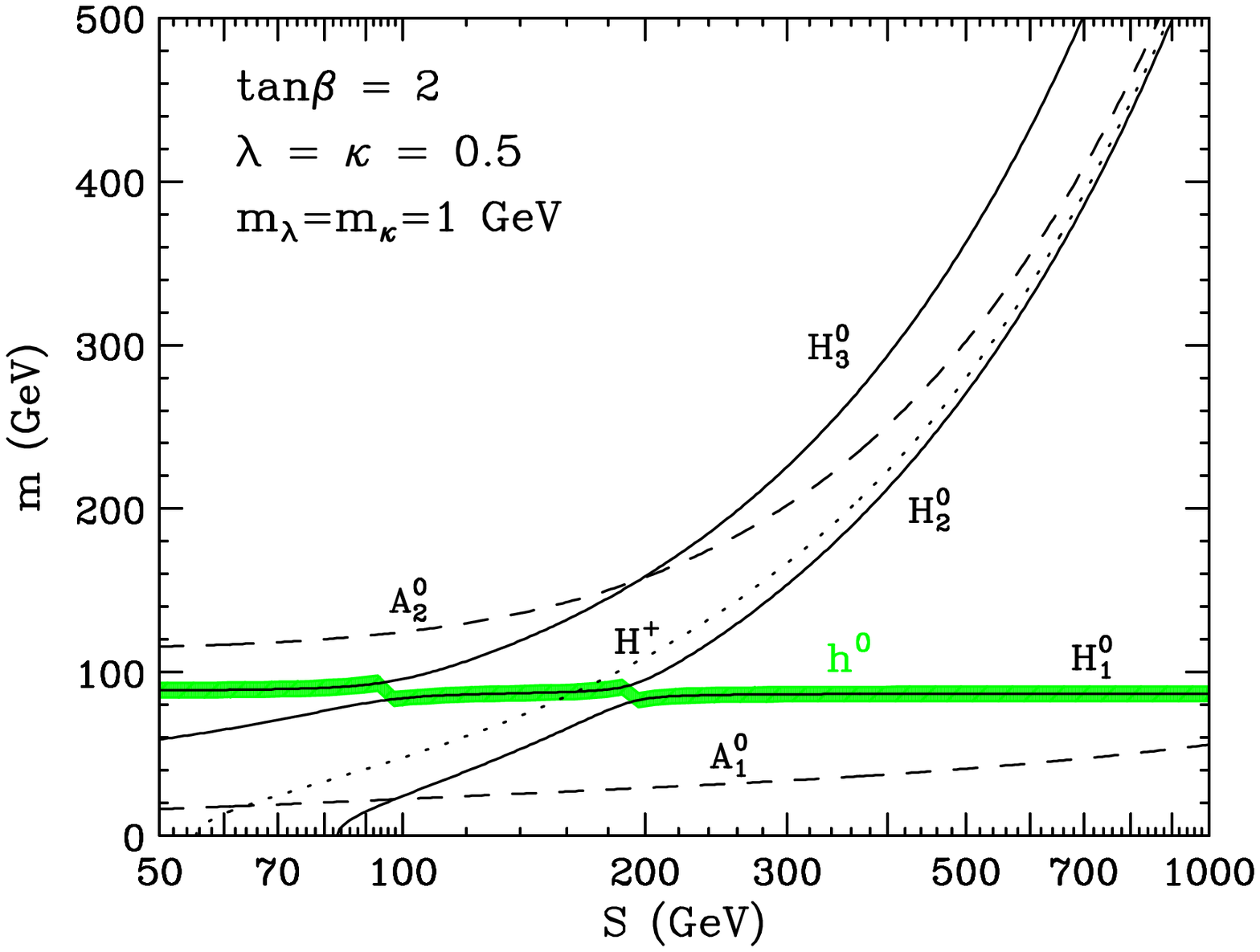,width=10cm} 
\caption{Higgs boson spectrum as a function of $s$,
for $\tan\beta=2$, $\lambda=\kappa=0.5$
and $m_\lambda=m_\kappa=1$ GeV.}%
\label{mhversuss}}
The above discussion is illustrated in Fig.~\ref{mhversuss},
where we plot the exact tree-level
masses of the CP-even Higgs bosons (solid lines),
the CP-odd Higgs bosons (dashed lines) and the charged Higgs boson
(dotted), as a function of $s$,
for fixed $\tan\beta=2$, $\lambda=\kappa=0.5$
and $m_\lambda=m_\kappa=1$ GeV. 
In order to guide the eye, we have added shading to trace
the SM-like Higgs boson $h^0$, which from right to left
is identified successively with $H^0_1$, $H^0_2$ and $H^0_3$.

In the limit $s\gg v$, the condition $M_{h^0}^2 > 0$
is necessary and sufficient for the existence 
of the electroweak asymmetric vacuum described by Eq.~(\ref{vacuum}).
The resulting constraint on the parameter space is given at tree-level by
\be
{\lambda^4\over2\kappa^2}\ <\
{M_Z^2\over v^2}\cos^2\! 2\beta
+ {\lambda^3\over\kappa}\, \sin2\beta.
\ee

Alternatively, for $s < v$, we see from Fig.~\ref{mhversuss}
that $H^0_1=H^0$ instead, and the requirement for positivity of
the Higgs masses squared implies a lower bound of $s$.
In any way, LEP-II bounds on $M_{h^0}$ should
provide a stronger constraint on the parameters than the 
requirement of a local minimum of the potential.

Just as an aside, notice that Eq.~(\ref{mhs}) obeys the
usual NMSSM tree-level upper bound on the lightest CP-even
scalar \cite{Drees:1989fc,Ellis:1989er},
\be
M_{H^0_1}^2 < M_Z^2 \cos^2\! 2\beta + {1\over2}\lambda^2v^2\sin^2\!2\beta~,
\ee
since Eq.~(\ref{mhs}) can be equivalently rewritten as
\be
M_{h^0}^2 = M_Z^2 \cos^2\! 2\beta + {1\over2}\lambda^2v^2\sin^2\!2\beta
        - {1\over2} \lambda^2 v^2
        \left(\sin2\beta-{\lambda\over\kappa}\right)^2.
\ee

In this decoupling limit, $h^0$ couples to the quarks and leptons exactly
like the Standard Model Higgs boson. Nevertheless, the decays of
$h^0$ may be very different than in the SM, if
$M_{A_1} < M_h/2$, since the $h^0 \rightarrow A^0_1A^0_1$
decay mode is then kinematically open.
In order to assess the partial width for this decay mode
\cite{Dobrescu:2000jt},
\begin{equation}
\Gamma(h^0\rightarrow A^0_1A^0_1)\ =\ 
{c^2\, v^2\over 32\pi M_{h^0}}\ 
\left(1-4{M_{A_1}^2\over M_{h^0}^2}  \right)^{1/2}~,
\label{width}
\end{equation}
we need to compute the coefficient $c$ of the trilinear term
${\cal L}_{h^0A_1^0A_1^0}$ shown in Eq.~(\ref{trilinear}).
Plugging the elements of $U$ shown in Eqs.~(\ref{us})
into Eq.~(\ref{haarot}), after a somewhat tedious calculation
we find a simple result:
\be
c = \frac{1}{2\kappa^2}
\left(\lambda - 2\kappa\sin 2\beta  \right) 
\left(\lambda +  \kappa\sin 2\beta  \right) \frac{M_h^2}{s^2}
+ {\cal O}\left({v^3\over s^3}\right)  ~.
\label{cs}
\ee
This compact formula has a simple physical interpretation.
In the polar coordinates parametrization of the axion field,
the axion has only derivative couplings, suppressed by the 
axion decay constant, $f_A$. Since the axion decay constant 
is given by the $U(1)_R$ breaking VEV, in the $s\gg v$ limit
we find $f_A \approx s$, while the derivatives in the coupling
yield a factor of $M_{h^0}^2$, after taking into account the
equations of motion. In addition, the coupling of the 
Higgs boson to pairs of axions has to be proportional to 
the Higgs VEV. Therefore, the amplitude for 
$h^0\rightarrow A^0_1A^0_1$ is proportional to $v M_{h^0}^2/s^2$.
The amplitude has to be the same in the 
polar and orthogonal coordinates, 
which explains both the appearance of the Higgs
boson mass $M_{h^0}$ rather than $v$ in the numerator of
Eq.~(\ref{cs}), as well as the cancellation of the terms of order
$(v/s)^0$ and $(v/s)^1$.
In the Appendix we provide the derivation of
the Higgs coupling to axion pairs 
in the polar coordinates parametrization.

Since $\lambda$ and $\kappa$ are expected to be of order one,
the size of $c$ is basically dictated by the degree of decoupling 
of the heavy scalars.
The quality of the approximation is shown in
Fig.~\ref{cversuss}, where we compare the prediction of the
analytical formula (\ref{cs}) (dashed line) with the exact numerical
results for the absolute value of the tree-level coefficient 
$|c|$ (solid lines),
as a function of $s$, for $\tan\beta=2$, $\lambda=\kappa=0.5$,
and three different values of $m_\lambda=m_\kappa$ (shown in GeV).
\FIGURE[t]{\epsfig{file=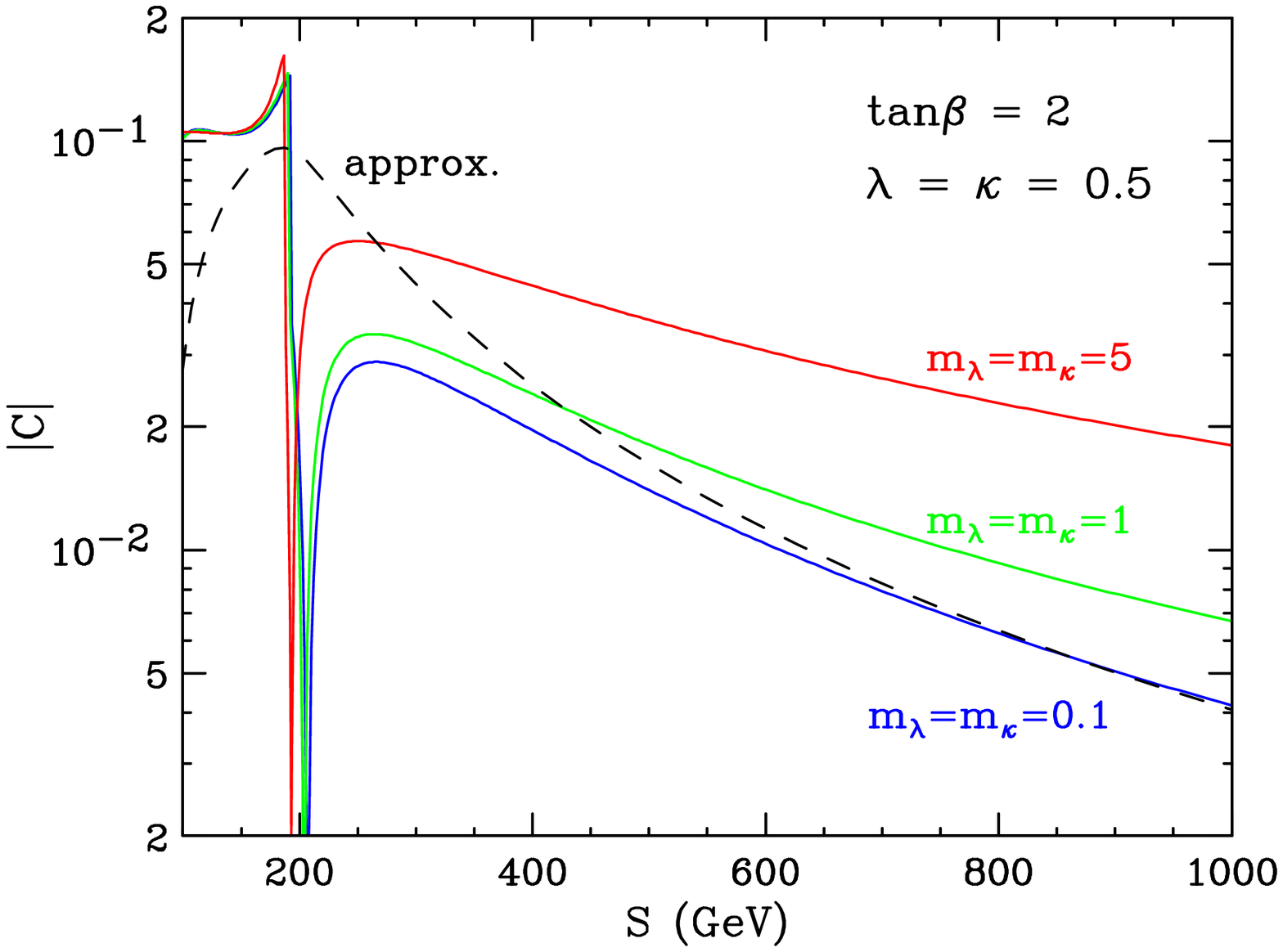,width=10cm} 
\caption{Comparison between the analytic approximation
(\ref{cs}) (dashed) and the exact results for $|c|$ (solid lines),
as a function of $s$, for $\tan\beta=2$, $\lambda=\kappa=0.5$,
and three different values of $m_\lambda=m_\kappa$ (shown, in GeV).}
\label{cversuss}}
We see that as expected, at large $s$ the analytic approximation
agrees pretty well with the exact result for the smallest
values of $m_\lambda$ and $m_\kappa$. The approximation fails
either at small $s$, or for larger $m_\lambda$ and $m_\kappa$
--- recall that we neglected terms of order $m_\lambda/v$ and
$m_\kappa/v$, while the leading term in Eq.~(\ref{cs})
is of order $v^2/s^2$. The dip around $s\sim 200$ GeV is due to
the fact that $c$ changes sign, as we switch from $h^0=H^0_1$
to $h^0=H^0_2$.

\section{The Large $\tan \beta$ Limit}
\label{largetb}

In this section we shall obtain the coefficient $c$
in the limit of large $\tan\beta$. To this end
we follow the procedure from the previous Section ---
expand $U$ in powers of $1/\tan\beta$ and substitute
the result in Eq.~(\ref{haarot}). However, we
will not take the limit $m_\lambda,m_\kappa\rightarrow0$,
but instead will keep the full dependence on $m_\lambda,
m_\kappa$, as this does not cause too much complication. 
Keeping only leading
order terms, Eq.~(\ref{haarot}) simplifies to
\be
c = - U_{i1}\, \lambda^2 
    + U_{i2}\, \lambda\kappa
    - U_{i3}\, 2\kappa^2\, {s\over v}~.
\label{c_largetb}
\ee
The relevant mixing angles in the CP-even Higgs sector
are also easy to compute --- notice that in the
large $\tan\beta$ limit, $H^0$ decouples and to leading
order $H^0_1$ is a linear combination of $h^0_v$ and $h^0_s$
only. We find 
\be
U_{11} = \cos\theta_H, \qquad
U_{12} = 0, \qquad
U_{13} = -\sin\theta_H,
\ee
where 
\be
\tan2\theta_H = {2\lambda^2 sv\over 2\kappa^2 s^2 - M_Z^2 - m_\kappa s}.
\ee
The coefficient $c$ is then simply
\be
c = - \lambda^2 \cos\theta_H
  + 2\kappa^2\, {s\over v}\, \sin\theta_H 
+ {\cal O}\left({1\over\tan\beta}\right) ~.
\label{ctb}
\ee

\FIGURE[t]{\epsfig{file=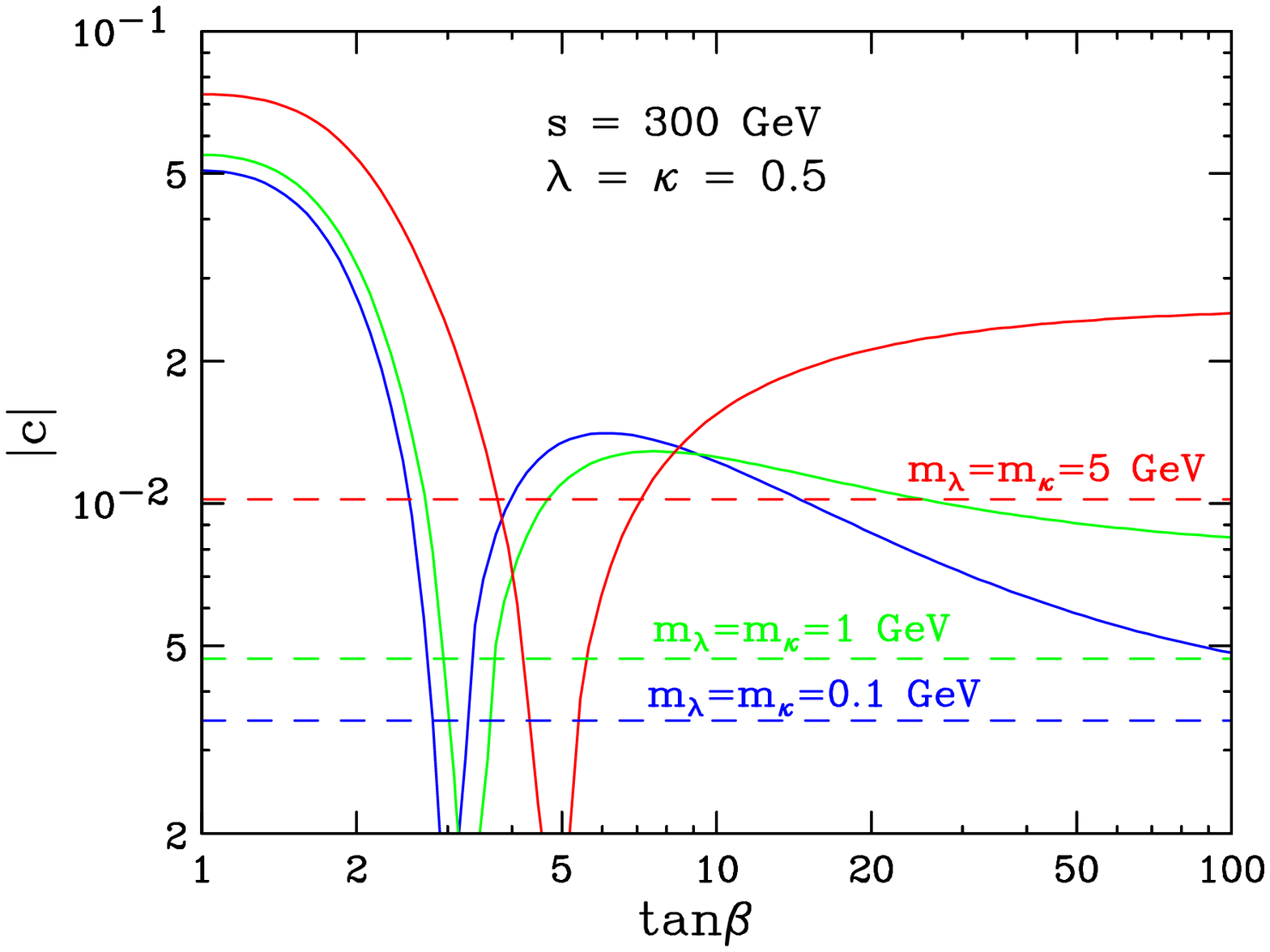,width=10cm} 
\caption{The same as Fig.~\ref{cversuss},
except plotted as a function of $\tan\beta$,
for fixed $s=300$ GeV. Unlike Fig.~\ref{cversuss},
here there is a different approximation for each value of
$m_\kappa$, as we kept the full $m_\kappa$ dependence 
in this section.}
\label{cversustb}}
In Fig.~\ref{cversustb} we show the validity of the
approximation (\ref{ctb}). In analogy to Fig.~\ref{cversuss},
we plot $|c|$, but this time versus $\tan\beta$, for a fixed
value of $s=300$ GeV. We see reasonable agreement between
the exact result and our approximation, given the
large cancellations between the leading order terms.
We have checked that each individual term contributing to
$c$ is reproduced with an accuracy of a few percent,
and the remaining difference in the total seen in the
Fig.~\ref{cversustb}
is due to a large cancellation between the first and third
terms of Eq.~(\ref{c_largetb}). Although the analytical 
approximations presented here are not intended to substitute the
full result, they illustrate these
leading order cancellations. It is also easy now to understand 
why $c$ vanishes for a particular value of $\tan\beta$ 
in Fig.~\ref{cversustb}: for $\sin2\beta = 
\lambda/(2\kappa)$ the analytical approximation (\ref{cs}) gives
$c=0$.

In conclusion of this section, we note that
when in addition we take the limit of large $s$,
Eq.~(\ref{ctb}) reduces to
\be
c = \frac{\lambda^2}{2\kappa^2} \frac{M_h^2}{s^2}
+ {\cal O}\left({v^3\over s^3},{1\over\tan\beta}\right) ~,
\label{cstb}
\ee
which is also in agreement with the large $\tan\beta$
limit of Eq.~(\ref{cs}).

\section{Numerical Results}
\label{numerical}

In this section we shall study numerically the scalar spectrum
and the coupling (\ref{trilinear}) of the Standard Model-like
Higgs boson $h^0$ to axion pairs.
There are several experimental (Section~\ref{subsec:exp}) and 
theoretical (Section~\ref{subsec:theory}) constraints on our scenario.
Most importantly, we must account for all existing
experimental bounds on the Higgs bosons and their superpartners.
(We do not consider experimental bounds from other
superpartner searches, since those depend on the particular
framework of supersymmetry breaking, which we never had to specify 
for our Higgs sector analysis.) 

The Higgs structure considered here must eventually be
embedded in some more fundamental theory, defined at a much higher
scale $\Lambda$, possibly the Planck or the string scale.
Hence, the low-energy Higgs sector parameters should be derived
in terms of the parameters from the fundamental theory without
excessive fine-tuning. It is also theoretically desirable 
that the couplings in the theory are free of Landau poles
at least up to the scale $\Lambda$. 

Our numerical analysis
in this Section is designed to address these issues.

\subsection{Collider constraints}
\label{subsec:exp}

Let us first start with the Higgs sector. LEP is typically
able to rule out new particles with order one couplings
close to its kinematic limit. For example, the charged Higgs bosons
$H^\pm$ can be produced in $s$-channel $Z/\gamma$ processes.
The current LEP bound is $m_{H^+}\gsim85$ GeV \cite{LEPHiggs} and we
expect it to be valid in the case of the NMSSM discussed here as well.

The Standard Model Higgs search at LEP, when reinterpreted as
a search for the lightest CP-even Higgs boson of the MSSM,
provides an additional constraint on our parameter space.
The current combined LEP limit is around 113 GeV.
As is well known however, the one-loop corrections to $M_{h}$
involving third generation quarks and their superpartners
are positive and potentially large \cite{Haber:1991aw},
so that the LEP constraint on the
{\em tree-level} Higgs boson mass $M_h$ is much weaker.
Depending on the stop/sbottom masses and the amount of squark mixing,
the radiative corrections can shift the tree-level value of
the Higgs mass by up to $\Delta M_h\sim 20-30$ GeV
\cite{Ellwanger:1993hn}.
Hence we shall allow for tree-level Higgs boson masses
as low as 80-90 GeV. Of course, within any given
supersymmetric model framework, one can compute this
difference exactly in terms of the parameters of the squark sector.
Here we prefer to stay within our model-independent approach
and avoid specifying a particular model of supersymmetry breaking and/or
squark spectrum and mixing angles, as this will hinder
the universal applicability of our results. Furthermore,
the novel effect of $h\rightarrow A^0_1A^0_1$ decays is maximally
operational for large $h^0$ masses (see, e.g. Eq.(\ref{cstb})),
where the LEP bound is likely to be satisfied.

Finally, LEP has also searched for superpartners of the $SU(2)$
Higgs bosons, as part of their chargino search.
Independent of their mass splitting, 
charged higgsinos lighter than $\sim 72$ GeV
\cite{grenier} have been ruled out. For typical values of
the higgsino mass splittings, the bound extends up to the
LEP kinematic limit. The higgsino masses are typically of order
$|\mu|$, where the $\mu$ parameter, familiar from the MSSM,
is given in our notation by
\be
\mu\ \equiv \ {\lambda s\over\sqrt{2}}~.
\ee
Thus the chargino mass limit constrains the product
of $\lambda$ and $s$. The exact bound depends on the amount
of mixing between the neutralinos, which
again would require us to specify the gaugino masses
within a particular model. It also depends on the particular
framework of supersymmetry breaking, {\it e.g.} 
in models with low-scale supersymmetry breaking
the bound from prompt decays of a higgsino NLSP
\cite{Baer:1999tx} from the Tevatron can be stronger.
We therefore again choose to stay on a model-independent path,
and we consider (rather conservatively) any value of
$\mu\gsim 100$ GeV as allowed.

Searches for the superpartners of the two neutral $SU(2)$-singlet
Higgs bosons ({\em i.e.} ``singlinos'') are quite challenging.
The couplings of the singlinos to gauge bosons are suppressed by
the neutralino mixing angles, and there are practically
no limits coming from direct singlino production or $Z$-decays
\cite{Franke:1996tf}.
The constraints on a light CP-odd scalar with small couplings
to SM fermions are also very loose 
\cite{Franke:1995xn,King:1996ys}. For example,
no significant constraints on the axion mass have been set from
$Z\rightarrow A^0\gamma$ at LEP \cite{ZAgamma},
from the Yukawa process $e^+e^-\rightarrow f\bar{f}A^0_1$
\cite{Djouadi:1991ms}, from direct $A^0$ production through
gluon fusion at the Tevatron \cite{Datta:1996qw},
from fits to the electroweak data \cite{Chankowski:1999ta},
or from meson decays \cite{HHG}.
The LEP searches for $Z^\ast\rightarrow h^0A^0\rightarrow A^0A^0A^0$
within the MSSM \cite{hAtoAAA} can be reinterpreted as 
axion searches, but the limits are diluted because the axion
coupling to the $Zh^0$ in our case is smaller by a factor of
$\cos\theta_A$ [see Eq.~(\ref{costhetaA})].
The relevant lower bounds on $M_A$ currently come from beam dump 
experiments \cite{Blumlein:1991ay}, in the MeV range, and from star 
cooling rates, $M_A \gsim 0.2$ MeV \cite{Groom:2000in}.

\FIGURE[ht]{\epsfig{file=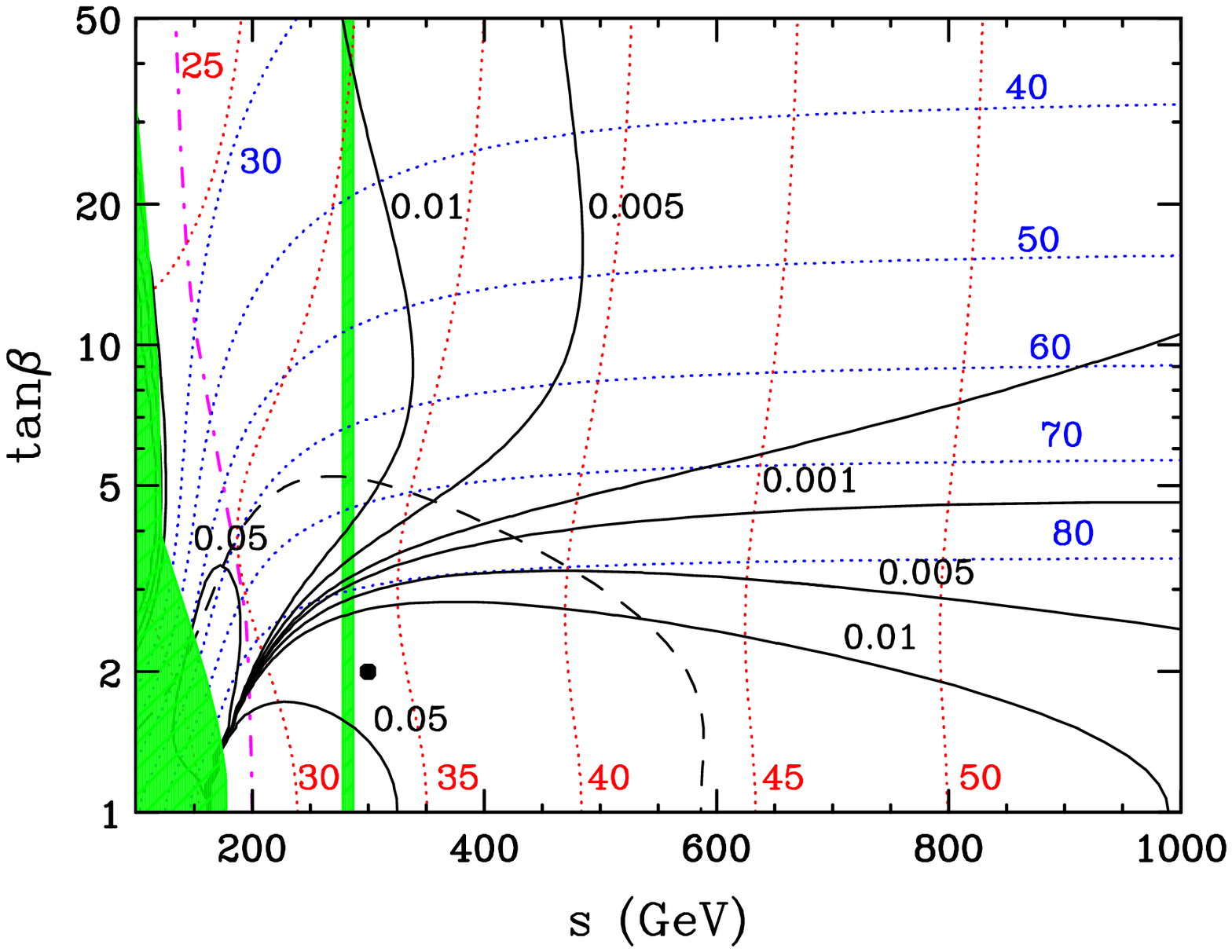,width=10cm} 
\caption{Higgs boson spectrum and the absolute
value of the dimensionless
$H^0_1 A^0_1 A^0_1$ coupling $c_{111}$ (see text),
as a function of $s$ (in GeV) and $\tan\beta$, for $\lambda=\kappa=0.5$
and $m_\lambda=m_\kappa=1$ GeV. The horizontal (blue)
dotted lines are contours of $M_{H^0_1}$, while
the vertical grid of dotted red lines consists of
$M_{A^0_1}$ iso-mass contours. The thick (green)
vertical line loosely marks the assumed higgsino mass
bound of $\mu=100$ GeV, while the (green) shaded area 
contains a tachyon or is excluded from the charged Higgs search at LEP.
The dot symbol denotes the values of $s$ and $\tan\beta$
used in Fig.~\ref{mhversuslk}. The dashed line delineates
the region where $M_{H^0_1}>2 M_{A^0_1}$ at tree-level.
To the right of the vertically running dotdashed (magenta)
line the SM-like CP-even Higgs boson
is identified with the lightest CP-even mass eigenstate: $h^0=H^0_1$,
while to the left of the line we have $h^0=H^0_2$ instead.
In the latter region, the corresponding coupling $c_{211}$ 
is larger, on the order of 10\% (see the small $s$ region of
Fig.~\ref{cversuss}).}
\label{mhversusst}}
Having summarized the relevant collider constraints, we
now present our exact numerical results for $c$.
In Fig.~\ref{mhversusst} we show contours of the masses of the
SM-like Higgs boson, $M_{h^0}$, [horizontal dotted (blue) lines]
and the lightest CP-odd Higgs boson, $M_{A^0_1}$, [vertical
dotted (red) lines] as a function of $s$ and $\tan\beta$,
for fixed $\lambda=\kappa=0.5$ and $m_\lambda=m_\kappa=1$ GeV.
The thick (green) vertical line denotes the assumed higgsino mass
bound of $\mu=100$ GeV and the region to its left is disfavored.
Inside the (green) shaded area there is either a tachyon in the
spectrum, or the charged Higgs mass is below the experimental bound.
While it is in principle possible that the radiative corrections
to the Higgs boson masses,               neglected here, may
shrink, or even eliminate this excluded region, we do not find
this latter part of parameter space particularly attractive, since 
it is associated with very light higgsinos.

Throughout most of the parameter space shown in Fig.~\ref{mhversusst},
the SM-like Higgs boson is the lightest CP-even mass eigenstate:
$h^0=H^0_1$. However, at small values of $s$, $H^0$ and $H'^0$ become
light as well (see Fig.~\ref{mhversuss}). The dotdashed (magenta) line
going vertically marks the point at which the SM-like Higgs boson
changes its identity --- from $h^0=H^0_1$ (to the right of the line) to
$h^0=H^0_2$ (to the left).

In Fig.~\ref{mhversusst} we only show the masses of the
$H^0_1$ and $A^0_1$ Higgs bosons, since we are interested in the
phenomenology of the $h^0\rightarrow A^0_1 A^0_1$ decay. 
The first important result seen in Fig.~\ref{mhversusst}
is the location of the region of parameter space,
where this decay is open. The dashed line in Fig.~\ref{mhversusst}
delineates the relevant part of parameter space where
$M_{H^0_1}>2 M_{A^0_1}$ at tree-level. The location
of this region is easily understood. Recall that $M_{A^0_1}$
scales with $\sqrt{s}$, as evident from the Figure, as well as
Eq.~(\ref{mA1}). Then
notice that unlike the case of the MSSM, here the lightest
CP-even Higgs boson mass {\em decreases} with $\tan\beta$,
so that $\tan\beta$ values as low as $1-3$ are possible.
Combining these two observations, we easily see that
the $h^0\rightarrow A^0_1 A^0_1$ decay is most likely to be
open at small $\tan\beta$ and small $s$.

Our second main result shown in Fig.~\ref{mhversusst} is the strength of the
Higgs to axion coupling. We show contours of the absolute value of the
dimensionless coupling $|c_{111}|$, which we 
define in analogy to Eq.~(\ref{trilinear}):
\be 
{\cal L}_{H^0_iA^0_jA^0_k} = \frac{v}{2}\
\sum_{i,j,k}\ c_{ijk} \ H^0_i A^0_jA^0_k ~.
\label{trilinearijk}
\ee
As we already explained above, to the right of the vertical dotdashed 
magenta line, $c_{111}$ is identical to the coefficient $c$ defined
in Eq.~(\ref{trilinear}). 
We see that in the region of parameter space, free
of any experimental constraints, $c$ can be as large as $0.05$.
We also see the possibility of exact cancellation
and vanishing $c$. Indeed, in the limit of large $s$, our
leading order approximation Eq.~(\ref{cs}) vanishes for
$\tan\beta=2+\sqrt{3}\sim 3.7$, in reasonable agreement with 
Fig.~\ref{mhversusst}. In the small $s$ region, where
$h^0=H^0_2$, the corresponding $c_{211}$ can be even larger,
on the order of 10\% or more.

In summary, our main conclusion from Fig.~\ref{mhversusst} is that
for fixed $\lambda$ and $\kappa$, values of $c$ 
are typically maximized at small $s$ and small $\tan\beta$ ---
exactly in the spot where the $h^0\rightarrow A^0_1 A^0_1$ channel
is most likely to be open. This could also have been anticipated from
the approximate scaling $c\sim M^2_{h^0}$, which we found
in the two limiting cases considered earlier.

\FIGURE[ht]{\epsfig{file=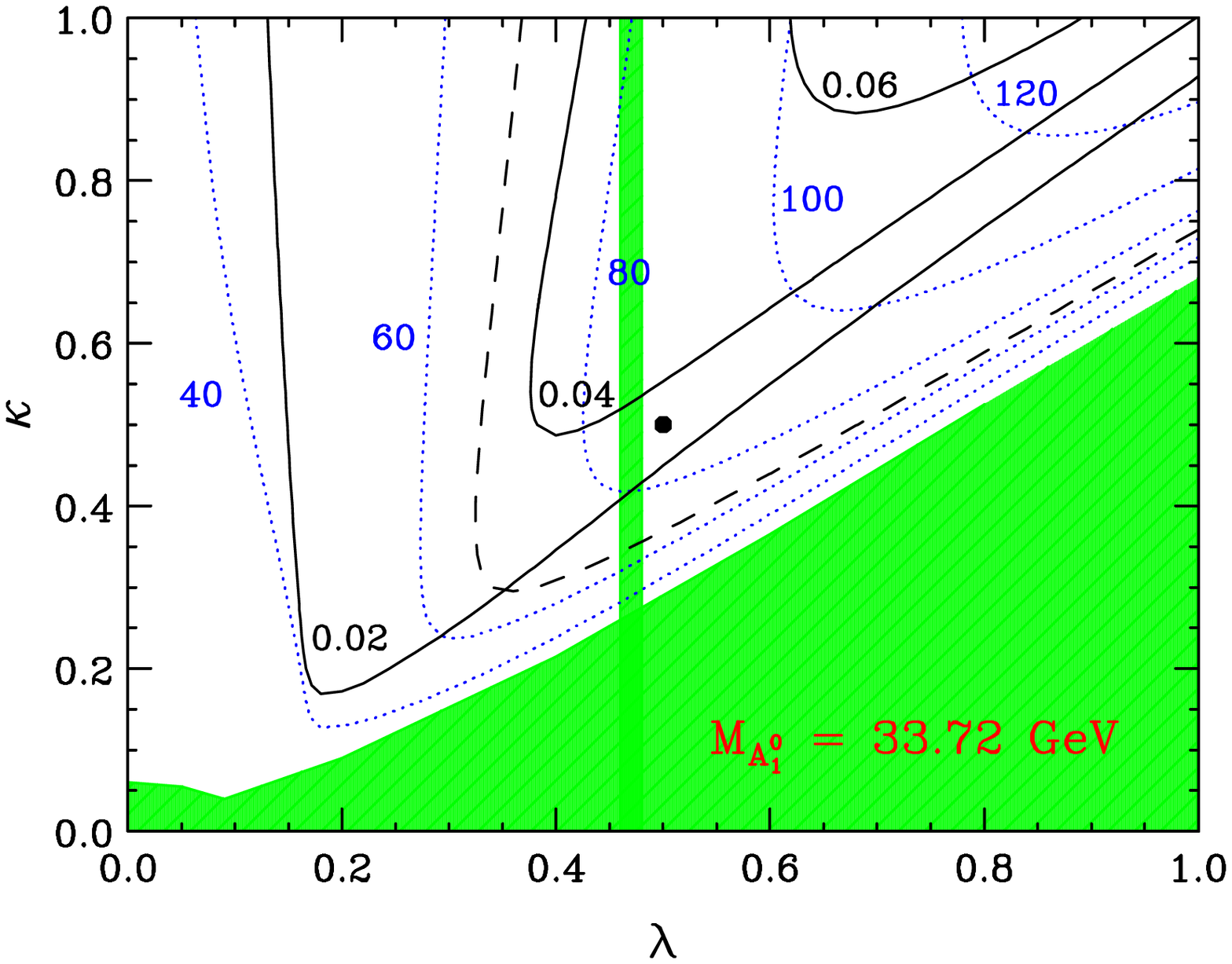,width=10cm} 
\caption{The same as Fig.~\ref{mhversusst}, but as a function of
$\lambda$ and $\kappa$, for $s=300$ GeV, $\tan\beta=2$
and $m_\lambda=m_\kappa=1$ GeV. The dot symbol shows the values
of $\lambda$ and $\kappa$ used in Fig.~\ref{mhversusst}.}
\label{mhversuslk}}
We now turn our attention to the dependence on the other two
main parameters: $\lambda$ and $\kappa$. To this end, we fix
$s=300$ GeV and $\tan\beta=2$, and show the Higgs spectrum
and $|c_{111}|$ in Fig.~\ref{mhversuslk} as a function of
$\lambda$ and $\kappa$, again with fixed $m_\lambda=m_\kappa=1$ GeV.
The (green) shaded area is again excluded because of tachyonic
or light charged Higgs states, and the (green) vertical line
denotes the assumed higgsino bound $\mu=100$ GeV, with the region
to the left of it being disfavored. This time the SM-like Higgs
boson is unambiguously identified as $h^0=H^0_1$.
On the other hand, the axion mass is completely fixed
in terms of $m_\lambda$, $m_\kappa$, $s$ and $\tan\beta$
[recall the discussion after Eq.~(\ref{tantheta})].
For the values of the parameters considered here, 
$M_{A^0_1}=33.72$ GeV and its fractional variation
throughout the whole Figure is less than 1 part in $10^{5}$.
The dashed line, depicting the region $M_{H^0_1}>2 M_{A^0_1}$,
is therefore coincident with the contour of $M_{H^0_1}=67.44$ GeV.

We see from Fig.~\ref{mhversuslk} that large ratios
of $\lambda/\kappa$ are disfavored. This can be easily understood
from Eq.~(\ref{mhs}) --- the second term gives a large negative
contribution to the lightest CP-even Higgs boson mass, which results
in $M^2_{h^0}<0$. Nevertheless, there is a significant allowed
region with naturally large couplings $\lambda$ and $\kappa$.
Now the lightest CP-even Higgs boson mass can be larger than
the LEP limit {\em already at tree-level}. 
The coefficient $c$ is again maximized in the region with
the largest $M_{h^0}$, and again values of $c\sim 0.05$ are possible.

Given that the small axion mass allows for the
$h^0\rightarrow A^0_1A^0_1$ decays in principle, 
and that the coefficient $c$ can be sizable
(see Figs.~\ref{mhversusst} and \ref{mhversuslk}),
it is interesting to quantify how the branching fractions
of the SM-like Higgs boson, $h^0$, are affected.
To this end, one must use the physical mass
$M_{h^0}^{\rm phys}$, as the partial widths for
$h^0\rightarrow A^0_1A^0_1$ and $h^0\rightarrow b\bar{b}$
have the opposite dependence on $M_{h^0}^{\rm phys}$.
So, by using the (smaller) tree-level mass, one would be 
overestimating $\Gamma(h^0\rightarrow A^0_1A^0_1)$
and underestimating $\Gamma(h^0\rightarrow b\bar{b})$.
\FIGURE[ht]{\epsfig{file=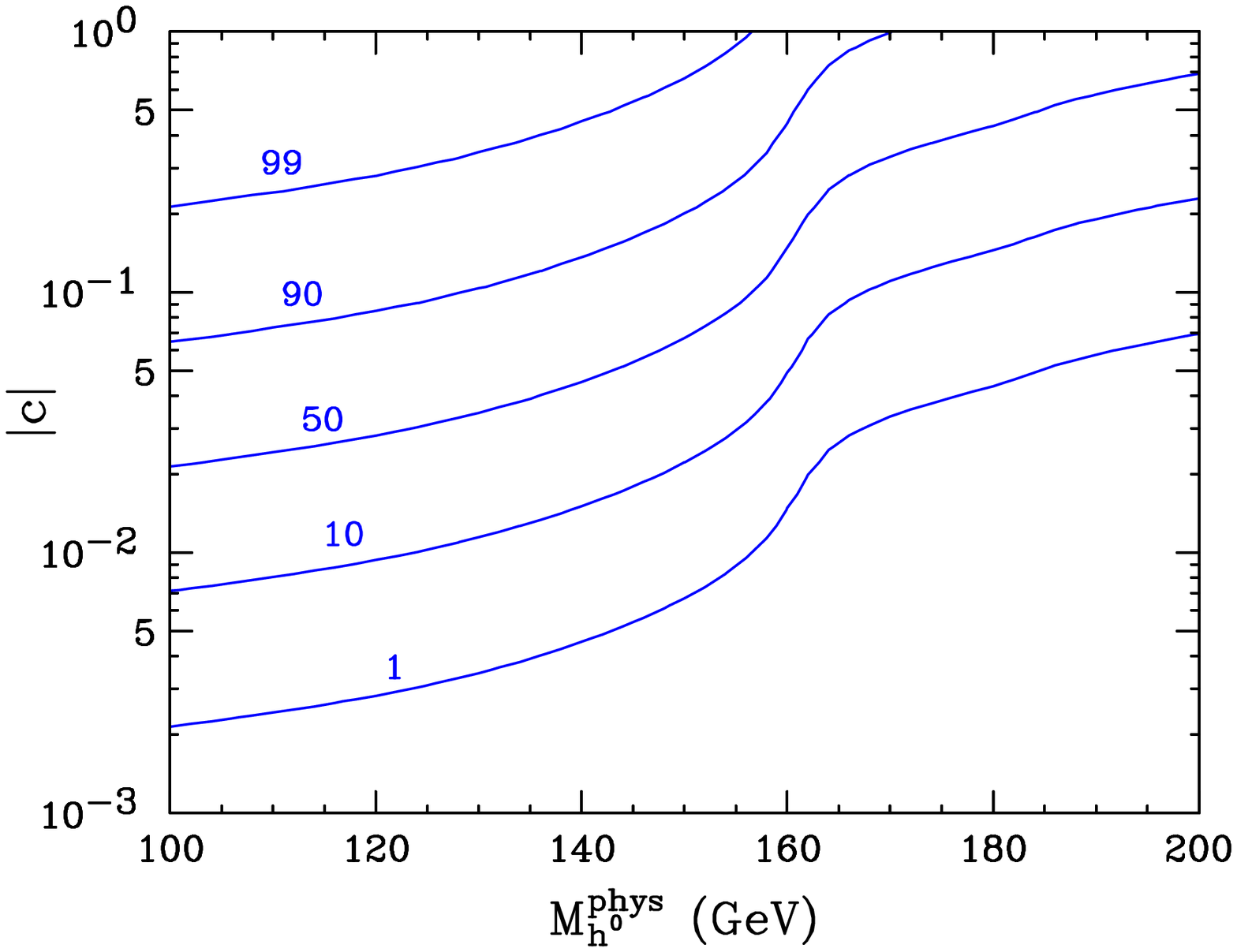,width=10cm} 
\caption{The branching ratio $B(h^0\rightarrow A^0_1A^0_1)$
in percent, as a function of the physical Higgs mass $M_{h^0}^{\rm phys}$
and the value of the coefficient $c$, assuming
$M_{A^0_1}\ll M_{h^0}^{\rm phys}$.}
\label{br}}
This is why in Fig.~\ref{br} we show the branching ratio for
$h^0\rightarrow A^0_1A^0_1$ decays
\be
B(h^0\rightarrow A^0_1A^0_1)\ =\
{\Gamma(h^0\rightarrow A^0_1A^0_1)\over
 \Gamma_{\rm SM} + \Gamma(h^0\rightarrow A^0_1A^0_1)}~,
\ee
as a function of the physical Higgs mass $M_{h^0}^{\rm phys}$
and the value of the coefficient $c$,
assuming $M_{A^0_1}\ll M_{h^0}^{\rm phys}$
and $M_{A^0_2}\gg M_{h^0}^{\rm phys}$. Here we have taken
$\Gamma_{\rm SM}$ as the width of the SM Higgs boson.
Fig.~\ref{br} should be interpreted as follows.
After fixing the fundamental input parameters
$\lambda$, $\kappa$, $m_\lambda$, $m_\kappa$, $s$ and $\tan\beta$,
one should compute the physical Higgs boson mass 
$M_{h^0}^{\rm phys}$ in terms of the
parameters of the squark sector derived within a given model.
Then the coefficient $c$ can be read off from
Figs.~\ref{mhversusst} and \ref{mhversuslk},
and the corresponding branching fraction
$B(h^0\rightarrow A^0_1A^0_1)$ for the resulting values
of $c$ and $M_{h^0}^{\rm phys}$ is given in Fig.~\ref{br}.

From Fig.~\ref{br} we can see the potential importance of the
$h^0\rightarrow A^0_1A^0_1$ decay mode. In the theoretically
preferred mass range $M_{h^0}^{\rm phys}\lsim 160-180$ GeV
of the NMSSM, this decay mode completely dominates for
$c\gsim 0.1$. We have seen that such values of $c$
are in principle possible, although in limited regions
of parameter space. For the more typical values of 
$c\sim 0.02-0.05$, the Higgs boson branching fraction into
axions is comparable, if not larger, than
$B(h\rightarrow b\bar{b})$, and may still significantly affect
the Higgs boson collider searches utilizing the $b\bar{b}$ mode.

\subsection{Theoretical prejudice }
\label{subsec:theory}

The theoretical constraints discussed in this subsection
are less robust, since one must define a framework.
For example, when requiring the absence of Landau poles
for the Higgs couplings $\lambda$ and $\kappa$, one must
specify the high-energy scale $\Lambda$, at which the effective theory
description of the NMSSM breaks down. Naturally, the constraints
would be stronger, if $\Lambda$ is identified with the GUT
or Planck scale, rather than some intermediate scale, related to
the (mediation of) supersymmetry breaking.

The Renormalization Group equations (RGE's) for the Yukawa
couplings of the NMSSM have been extensively studied in relation
to the question of the absolute upper limit on the lightest CP-even Higgs
mass \cite{terVeldhuis:1992es,Ellwanger:1993jp}.
If $\Lambda$ is identified with the GUT scale,
the perturbativity requirement implies $\lambda\lsim 0.87$
and $\kappa\lsim 0.63$
\cite{Ellis:1989er,Espinosa:1992gr,terVeldhuis:1992es,King:1996ys}.
If, however, $\Lambda$ is identified with an intermediate scale,
as in gauge-mediated models \cite{Han:2000jc,deGouvea:1998cx},
then the bounds can be somewhat relaxed, {\it e.g.}
$\kappa\lsim 1.36$ for $\Lambda\sim 50-100$ TeV \cite{Han:2000jc}.
We see that these constraints still leave a lot of the
available parameter space in Figs.~\ref{mhversusst}
and \ref{mhversuslk}, where the new decay $h^0\rightarrow A^0_1A^0_1$
is possible, and its branching ratio is sizable.

One may also wonder if the low values
of $m_\lambda$ and $m_\kappa$ needed in order to
suppress the axion mass $M_{A_1}$ (see Eq.~(\ref{mA1}))
can appear naturally, without any significant fine-tuning.
So far we have adopted a low-energy point of view and never
specified the particular model framework for supersymmetry breaking
and its communication to the NMSSM sector.
The size of $m_\lambda$ and $m_\kappa$ will depend
on two factors: the boundary conditions at the
high-energy scale $\Lambda$, where the soft supersymmetry breaking
parameters are generated, and second, the amount of
(logarithmic) RGE running from that scale
down to the electroweak scale $v$.
Notice that due to the singlet nature of $\hat{S}$,
the one-loop beta function for
$m_\kappa$ only depends on $m_\lambda$ and $m_\kappa$.
In the $U(1)_R$ limit both of these two parameters
start out small at the high-energy scale $\Lambda$,
and the generated value for $m_\kappa$ at the weak scale
is typically also rather small. On the other hand,
the one-loop beta function for $m_\kappa$ depends on the
gaugino masses $M_1$ and $M_2$, as well as the rest of
the $A$-term parameters, which can be relatively large
already at the scale $\Lambda$. Thus the induced
value for $m_\lambda$ at the weak scale can be much larger
than $m_\kappa$. However, a closer inspection of Eq.~(\ref{mA1})
reveals that the axion mass is usually more sensitive to
$m_\kappa$ rather than $m_\lambda$. For example, in the 
large $\tan\beta$ limit we have $\cos\theta_A\sim 1/\tan\beta$,
hence
$$
M^2_{A^0_1}\ \sim\ {\cal O}\left(m_\kappa s\right) + 
{\cal O}\left({m_\lambda s\over \tan\beta}\right)\
\sim\ {\cal O}\left(m_\kappa s\right).
$$
Without a particular model, it is difficult
to be more specific at this point. We shall therefore
leave this question open for future studies.
Let us only point out that there exist at least two
very well motivated frameworks, in which the
boundary conditions for $m_\lambda$ and $m_\kappa$
are zero at the scale $\Lambda$ --- gauge mediation
\cite{Dine:1993yw,Dimopoulos:1997yq}
and gaugino mediation \cite{Kaplan:1999ac}.

\section{Conclusions}
\label{conclusions}

We have discussed a limit of the NMSSM where a light
CP-odd scalar (axion) is present in the Higgs spectrum.
The light axion appears as a result of an approximate global
$U(1)_R$ symmetry of the scalar potential, which
is  spontaneously broken. We found that the mass of the axion is
proportional to the soft breaking trilinear
couplings $m_\lambda$ and $m_\kappa$, and if those are in the
GeV range, the axion can easily be lighter than
half the SM-like Higgs boson mass, $M_h^0$.
In those cases, we computed the $h^0A^0_1A^0_1$
coupling and found that it can have a direct impact
on phenomenology, as it can substantially modify the
SM-like Higgs boson collider signatures.

In conclusion, we feel that our results fill a major gap
in the extensive literature on the Higgs sector of the NMSSM.
Previous NMSSM studies have mostly concentrated on setting
absolute upper limits on the SM-like Higgs boson mass \cite{lump}
(which is of primary interest for the production of $h^0$
in collider experiments) or the related singlino phenomenology
\cite{Franke:1996tf,singlino}. However, the case of a light
CP-odd axion considered here has largely been overlooked.
In light of the interesting phenomenological implications
of the scenario presented here, and the symmetry reasons behind 
its motivation, it is worth pursuing the case of a light CP-odd
scalar in future phenomenological studies.

\bigskip

\acknowledgments{We thank M.~Carena, H.-C.~Cheng, S.~Mrenna,
A.~Pilaftsis, M.~Schmaltz and C.~Wagner for comments and questions.
Fermilab is operated under DOE contract DE-AC02-76CH03000.}

\section*{Appendix: \ Higgs Coupling to Axions in Polar Field Coordinates}
\renewcommand{\theequation}{A.\arabic{equation}}
\setcounter{equation}{0}

In this Appendix we derive the Higgs boson coupling to axion pairs 
using the polar field coordinates. Although the physical results 
are independent of the parametrization of the degrees of freedom,
and the orthogonal field coordinates used in section 2 [see
Eq.~(\ref{vacuum})] are convenient enough, the polar field coordinates
are particularly appropriate for describing the axion in the limit
where we neglect its mass.

Consider the following parametrization:
\bear
H_d & = & \left( \begin{array}{c}
\frac{1}{\sqrt{2}} 
\left[ (v + h^0_v )\cos\beta 
     - H^0_v \sin\beta \right] e^{i(A^0_v \tan\beta - G^0)/v} \\ [4mm]
                     -\ G^- \cos\beta 
                      + H^- \sin\beta 
\end{array} \right) \; ~,
\nonumber \\ [7mm]
H_u & = &\left( \begin{array}{c} 
                        G^+ \sin\beta + H^+ \cos\beta  \\ [4mm]          
      \frac{1}{\sqrt{2}}
\left[ (v + h^0_v)\sin\beta + H^0_v\cos\beta \right] 
e^{i(A^0_v \cot\beta + G^0)/v}
\end{array} \right) \; ~,
\nonumber \\ [6mm]
S & = & \frac{1}{\sqrt{2}} \left( s + h^0_s\right)e^{i A^0_s/s} \; ~.
\label{vacuum-polar}
\eear
With this parametrization, it is straightforward to derive 
the masses and mixing angle of the CP-odd states obtained in section
3.2. 

More importantly, 
the scalar potential Eq.~(\ref{vacuum}) does not involve the axion field 
in the $m_{\lambda,\kappa}\rightarrow 0$ limit. 
If we neglect the axion mass, then only the kinetic terms give rise to 
axion couplings. This is the usual statement that a Nambu-Goldstone 
boson has derivative couplings in the polar field coordinate
parametrization.
The kinetic terms for $H_u, H_d$ and $S$ include 
canonically normalized kinetic terms for the physical states as well
as derivative couplings of CP-odd scalar pairs with CP-even scalars.
The SM-like Higgs boson coupling to axion pairs can be easily derived:
\be
{\cal L}^\prime_{h^0A_1^0A_1^0} = \frac{c^\prime v}{2 M_h^2}  \ h^0
\partial_\mu A_1^0 \partial^\mu A_1^0 ~,
\label{trilinear-polar}
\ee
where the dimensionless parameter $c^\prime$ is given by 
\be
c^\prime = 2 \frac{M_h^2}{v^2} \left[ \left(U_{11} + 2U_{12} \cot 2\beta
\right)\cos^2\!\theta_A + \frac{v}{s} U_{13} \sin^2\!\theta_A \right] ~.
\ee
In the $s \gg v$ limit, the elements of the 
matrix $U$ that rotates the CP-even
scalars to the mass eigenbasis have been computed in section 4. 
Using Eq.~(\ref{us}) we find a very simple result:
\be 
c^\prime = -2c\left[ 1 + {\cal O}(v/s, m_{\lambda,\kappa}/s)\right] ~,
\ee
where $c$ is given by Eq.~(\ref{cs}).  
The last step of this computation is to check that the 
decay width for $h^0 \rightarrow A_1^0 A_1^0$ is the same as the one
obtained using orthogonal coordinates [see Eq.~(\ref{width})].
This straigthforward exercise  provides the 
explanation for the $M_h^2/s^2$ dependence of $c$ in Eq.~(\ref{cs}).  

\listoftables		
\listoffigures		


\end{document}